%% file: main.tex
\pdfoutput=1

\documentclass[11pt]{article}

\usepackage[]{acl}

\usepackage{times}
\usepackage{latexsym}

\usepackage[T1]{fontenc}

\usepackage[utf8]{inputenc}

\usepackage{microtype}

\usepackage{times}
\usepackage{soul}
\usepackage{cancel}
\usepackage{url}
\usepackage{hyperref}
\usepackage{graphicx}
\usepackage{amsmath}
\usepackage{amsthm}
\usepackage{booktabs}
\usepackage{algorithm}
\usepackage{algorithmic}
\usepackage{xcolor}         
\usepackage{bm}
\usepackage{amsfonts}

\newcommand{\report}[2]{${#1}^{#2}$}

\usepackage{xparse}
\NewDocumentCommand{\heng}
{ mO{} }{\textcolor{red}{\textsuperscript{\textit{Heng}}\textsf{\textbf{\small[#1]}}}}
\NewDocumentCommand{\manling}
{ mO{} }{\textcolor{blue}{\textsuperscript{\textit{Manling}}\textsf{\textbf{\small[#1]}}}}
\NewDocumentCommand{\zhenhailong}
{ mO{} }{\textcolor{orange}{\textsuperscript{\textit{Zhenhailong}}\textsf{\textbf{\small[#1]}}}}
\NewDocumentCommand{\francis}
{ mO{} }{\textcolor{green}{\textsuperscript{\textit{francis}}\textsf{\textbf{\small[#1]}}}}
\NewDocumentCommand{\han}
{ mO{} }{\textcolor{magenta}{\textsuperscript{\textit{Han}}\textsf{\textbf{\small[#1]}}}}

\title{Rethinking Task Sampling \\ for Few-shot Vision-Language Transfer Learning}

\author{
Zhenhailong Wang,
Hang Yu,
Manling Li,
Han Zhao,
Heng Ji\\
University of Illinois at Urbana-Champaign\\
\{wangz3, hengji\}@illinois.edu
}

\begin{document}

\maketitle

\begin{abstract}
Despite achieving state-of-the-art zero-shot performance, existing vision-language models still fall short of few-shot transfer ability on domain-specific problems. Classical fine-tuning 
often fails to prevent highly expressive models from exploiting spurious correlations 
in the training data. 
Although model-agnostic meta-learning (MAML) presents as a natural alternative for few-shot transfer learning, the expensive computation due to implicit second-order optimization limits its use on large-scale vision-language models such as CLIP. 
While much literature has been devoted to exploring alternative optimization strategies, we identify another essential aspect towards effective few-shot transfer learning, \emph{task sampling}, which is previously only be viewed as part of data pre-processing in MAML. 
To show the impact of task sampling, we propose a simple algorithm, Model-Agnostic Multitask Fine-tuning (MAMF), which differentiates classical fine-tuning only on uniformly sampling multiple tasks. Despite its simplicity, we show that MAMF consistently outperforms classical fine-tuning on five few-shot image classification tasks. We further show that the effectiveness of the bi-level optimization in MAML is highly sensitive to the zero-shot performance of a task in the context of few-shot vision-language classification. The goal of this paper is to provide new insights on what makes few-shot learning work, and encourage more research into investigating better task sampling strategies. Code and processed data are publicly available for research purposes at \url{https://github.com/MikeWangWZHL/Multitask-Finetuning_CLIP}
\end{abstract}

\input{sections/introduction}
\input{sections/method}
\input{sections/experiment}

\input{sections/conclusion}

\bibliography{anthology,custom}
\bibliographystyle{acl_natbib}

\appendix

\input{sections/appendix}

\end{document}

%% file: sections/introduction.tex
\begin{figure*}[th!]
  \centering
  \includegraphics[width=0.85\textwidth]{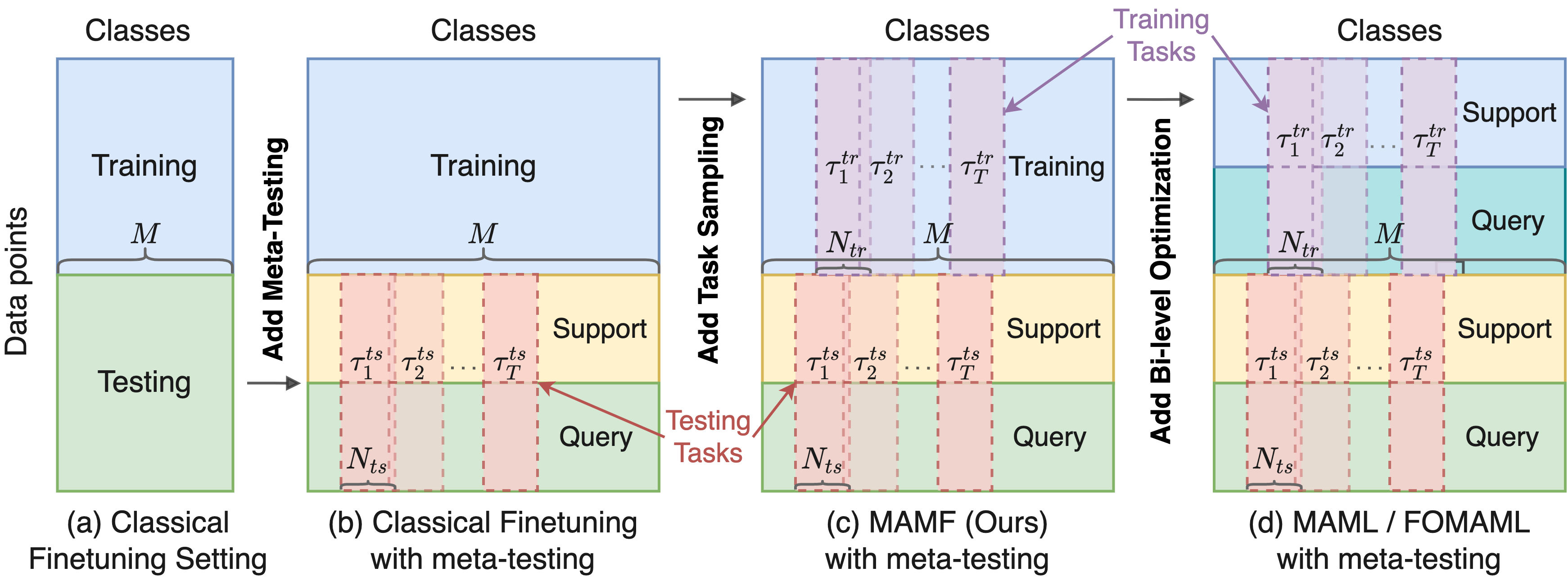}
  \caption{Task sampling and optimization schemes of different algorithms. Evaluation with meta-testing is applied in all of our experiments (b,c,d).
}
  \label{fig:algorithm}
\end{figure*}

\section{Introduction}
\label{sec:intro}


While existing machine learning models have achieved human-level performance at various individual tasks, they generally lack the ability of fast adaptation and generalization. In recent years, transfer learning has been proven to be effective on a wide range of Computer Vision~\cite{he2016deep,dosovitskiy2020image} and Natural Language Processing~\cite{devlin-etal-2019-bert,lewis-etal-2020-bart} tasks. Specifically, recent advances in large-scale vision-language models~\cite{pmlr-v139-radford21a,jia2021scaling,li2022blip,alayrac2022flamingo} have demonstrated strong zero-shot ability on a wide range of tasks. However, these models still have certain limitations on concepts that require extensive domain knowledge, such as Fungi Classification. We identify two major limitations in current few-shot transfer learning literature, from both evaluation and algorithm perspective.


\noindent\textbf{Limitation on evaluation}\;\; 
In current transfer learning paradigm, the testing instances of a downstream task are drawn from the same distribution as the training set. This evaluation setting can fail to faithfully reflect whether a model has truly learned a new concept, since modern deep neural networks can easily memorize and exploit spurious correlations from the training set~\cite{brown2020language}.
Thus, we first propose a new evaluation scheme for \textit{few-shot transfer learning} where we replace the original testing phase with \emph{meta-testing} (Section~\ref{sec:reformulate}). With \emph{meta-testing}, the testing distribution are distinguished from the training.

\noindent\textbf{Limitation on algorithm}\;\;
To make an arbitrary pretrained vision-language model learn new concepts with few examples, model-agnostic meta-learning (MAML)~\cite{finn2017model} presents as a natural candidate. One major limitation of the original MAML method is the expensive computation overhead due to implicit second-order optimization. Most follow-up work~\cite{finn2017model,nichol2018first, rajeswaran2019meta,Raghu2020Rapid,von2021learning} has focused on improving the optimization strategy. However, we found that they all achieved comparable performance despite of using different optimization algorithms. This observation motivates us to ask: \emph{If the specific choice of optimization method is not the key to the empirical success of MAML, what would be?}

Inspired by related work in the area of multitask learning~\cite{maurer2016benefit,tripuraneni2020theory}, 
we conjecture that \emph{task sampling} itself is an essential ingredient in learning new concepts efficiently.
To verify this hypothesis, we propose a simple fine-tuning algorithm, \emph{Model-Agnostic Multitask Fine-tuning (MAMF)}, which simplifies MAML by using only first-order gradient-based optimization while keeping the uniform task sampling procedure intact. The goal is \textbf{NOT} to propose yet another complex algorithm, but to investigate what is the most important aspect for effective few-shot transfer learning.
We compare MAMF with Classical Fine-tuning, which does not perform uniform task sampling, and first-order MAML (FOMAML)~\cite{finn2017model}, which adopts complex bi-level optimization upon sampled tasks.
Our empirical result demonstrates the importance of \textbf{\textit{uniform task sampling}} and reveals limited effectiveness of the bi-level optimization of MAML in the context of few-shot transfer learning.
We hope our work encourages more research into exploring better task sampling strategies for improving few-shot transfer learning and meta-learning algorithms.

%% file: sections/method.tex
\section{Problem Formulation} 
\label{sec:problem_formulation}
We are interested in a few-shot classification problem where we have a pretrained vision-language model $f$ with initial parameters $\bm{\theta}$. Let $\tau^{tr}$ be a training task sampled from a distribution $p(\tau^{tr})$, and $\tau^{ts}$ be a testing task sampled from $p(\tau^{ts})$, where a \textbf{task} is defined to an \textbf{induced sub-problem by restricting the output space from the original problem}. Specifically, for an original classification problem with $M$ classes in total, we define a task as a sub-problem where the output space is a subset of $N$ classes randomly sampled from the $M$ classes. We further denote $N^{tr}$ and $N^{ts}$ as the number of classes in each training and testing task. $T^{tr}$ and $T^{ts}$ as the total number of sampled tasks respectively.  The Classical Fine-tuning setting is depicted in Figure~\ref{fig:algorithm} (a), where we have $T^{tr} = 1$ training tasks with $N^{tr} = M$ classes, and $T^{ts} = 1$ testing tasks with $N^{ts} = M$ classes. That is, both training and testing sets are treated as one single task containing data points from all $M$ classes.  



\section{Reformulating Classical Fine-tuning Evaluation with Meta-testing}
\label{sec:reformulate}


Our goal is to enable and evaluate a model's capability of generalizing to new concepts with few examples. 
The Classical Fine-tuning setting is not sufficient since the training and testing data points are drawn from the same distribution. 
Therefore, we propose to replace the original joint testing in Classical Fine-tuning with \emph{meta-testing}.

Meta-testing is first introduced by related work in meta-learning
~\cite{thrun2012learning,vinyals2016matching,finn2017model}. As shown in the testing phase of Figure~\ref{fig:algorithm} (b,c,d), we first sample $T^{ts}$ tasks ($T^{ts}>1$), each containing data points from $N^{ts}$ classes ($1<N^{ts}<M$). For each sampled testing task $\tau^{ts}$, we further randomly split the data points into two disjoint sets, i.e., support set $A$ and query set $B$, with corresponding loss $\mathcal{L}_{\tau^{ts},A}$ and $\mathcal{L}_{\tau^{ts},B}$. Then we further update the model parameters on the support set and evaluate on the query set. By randomly sampling multiple tasks during \emph{meta-testing}, we can distinguish the testing distribution from training, which largely prevents the model from exploiting spurious correlations in the training set. Essentially, we make the original problem more challenging by requiring the model to quickly generalize to potentially unseen task distributions during testing. The objective is to find an updated model parameter $\widetilde{\bm{\theta}}$ that minimizes the expected loss on all testing tasks $\mathbb{E}_{\tau^{ts}\sim p(\tau^{ts})}\left[\mathcal{L}_{\tau^{ts}}(\widetilde{\bm{\theta}})\right]$.
Specifically, under this setting, MAML's objective can be written as follows:
\begin{align*}
    &\min_{\widetilde{\bm{\theta}}}\mathbb{E}_{\tau^{ts}\sim p(\tau^{ts})}\left[ \mathcal{L}_{\tau^{ts},B}\left(U_{\tau^{ts},A}^{ts}(\widetilde{\bm{\theta}}) \right)\right],\\
    &\;\;\widetilde{\bm{\theta}} = \min_{\bm{\theta}}\mathbb{E}_{\tau^{tr}\sim p(\tau^{tr})}\left[ \mathcal{L}_{\tau^{tr},B}\left(U_{\tau^{tr},A}^{tr}(\bm{\theta}) \right)\right]
\end{align*}
where $U_{\tau^{tr},A}^{tr}$ is the optimization procedure that updates the initial parameter $\bm{\theta}$ for one or more steps on the support set of a training task $\tau^{tr}$.

\section{Model-Agnostic Multitask Fine-tuning}


As shown above, previous MAML-like methods update model parameters iteratively via a complex bi-level optimization scheme~\cite{finn2017model,Raghu2020Rapid,rajeswaran2019meta}, which is computationally expensive.
We hypothesize that the \textit{task sampling} process itself is more important than specific choice of optimization method. 
To verify this hypothesis, we propose a simple algorithm, \emph{Model-Agnostic Multitask Fine-tuning (MAMF)}, where we keep the uniform task sampling strategy as MAML but perform simple first-order gradient-based optimization on each task sequentially. Unlike MAML, MAMF does not further split the tasks into support and query sets. 
The objective of MAMF can be written as:
\begin{align*}
    &\min_{\widetilde{\bm{\theta}}}\mathbb{E}_{\tau^{ts}\sim p(\tau^{ts})}\left[ \mathcal{L}_{\tau^{ts},B}\left(U_{\tau^{ts},A}^{ts}(\widetilde{\bm{\theta}}) \right)\right]\\
    &\widetilde{\bm{\theta}} = \bm{\theta}_{i=T^{tr}}, \bm{\theta}_{i}=U^{tr}_{\tau^{tr}_i}(\bm{\theta}_{i-1}), i \in \{1,2,...,T^{tr}\}
\end{align*}
where $\bm{\theta}_{0} = \bm{\theta}$ and $U^{tr}_{\tau^{tr}_i}$ is the optimization procedure that updates the parameters from the previous task on the current training task $\tau^{tr}_i$. MAMF can also be viewed as a simplified version of Reptile~\cite{nichol2018first}, where we further eliminate the hyper-parameter of step size. The goal is to keep the algorithm as simple as possible to distinguish the impact of \textit{task sampling}.
Figure~\ref{fig:algorithm} depicts a comparison of different data sampling and optimization schemes of different algorithms. 

%% file: sections/experiment.tex
\section{Experiment}

\subsection{Experimental Setup}
We aim to investigate \textbf{two main questions} experimentally under a \textbf{few-shot vision-language transfer learning setting}: 
\vspace{-4pt}
\begin{itemize}
    \item \textbf{Q1}: Is the \emph{uniform task sampling} during training important?  
    \vspace{-8pt}
    \item \textbf{Q2}: Is the \emph{bi-level optimization} in MAML consistently effective?
\end{itemize}
\vspace{-6pt}
To answer the first question, we compare MAMF with Classical Fine-tuning where the only difference is the additional uniform task sampling. For the second question, we compare FOMAML\footnote{We use the first-order variant of MAML for apple-to-apple comparison with MAMF.}
and MAMF. 


We perform comprehensive experiments on five few-shot image-classification datasets with various domains, including ClevrCounting~\cite{johnson2017clevr}, Amazon Berkeley Objects (ABO)~\cite{collins2021abo} Material, Fungi~\cite{su2021realistic}, Mini-Imagenet~\cite{vinyals2016matching}, Caltech-UCSD Birds 200 (CUB)~\cite{WelinderEtal2010}. We compare different learning algorithms by applying them to a large-scale vision-language model, i.e., CLIP~\cite{pmlr-v139-radford21a}. We adopt the contrastive classification framework following \cite{pmlr-v139-radford21a} where we directly match prompted label text with encoded images. This framework allows us to avoid the label permutation problem raised by \cite{ye2021train}. Details on the datasets and the classification framework can be found in Appendix~\ref{sec:appendix:dataset} and ~\ref{sec:appendix:classification_framework}. 

Given a dataset with $M$ classes in total, we experiment with various task configurations regarding the number of sub-sampled classes $N^{ts}$, where $2 \leq N^{ts} \leq M$. That is, during \emph{meta-testing}, each task can be formulated as a $N^{ts}$-way classification and we randomly sample $T^{ts}$ such tasks.
During training, for Classical Fine-tuning, we set the training task configuration as $N^{tr} = M, T^{tr} = 1$; for MAMF and FOMAML, we set $N^{tr} = N^{ts} = N, T^{tr}=T^{ts}=T$, where $T$ is determined based on $N$ to cover all classes with a high probability. Implementation details can be found in Appendix~\ref{sec:appendix:implementation_details}. 

\begin{figure*}[th]
  \centering
  \includegraphics[width=0.87\textwidth]{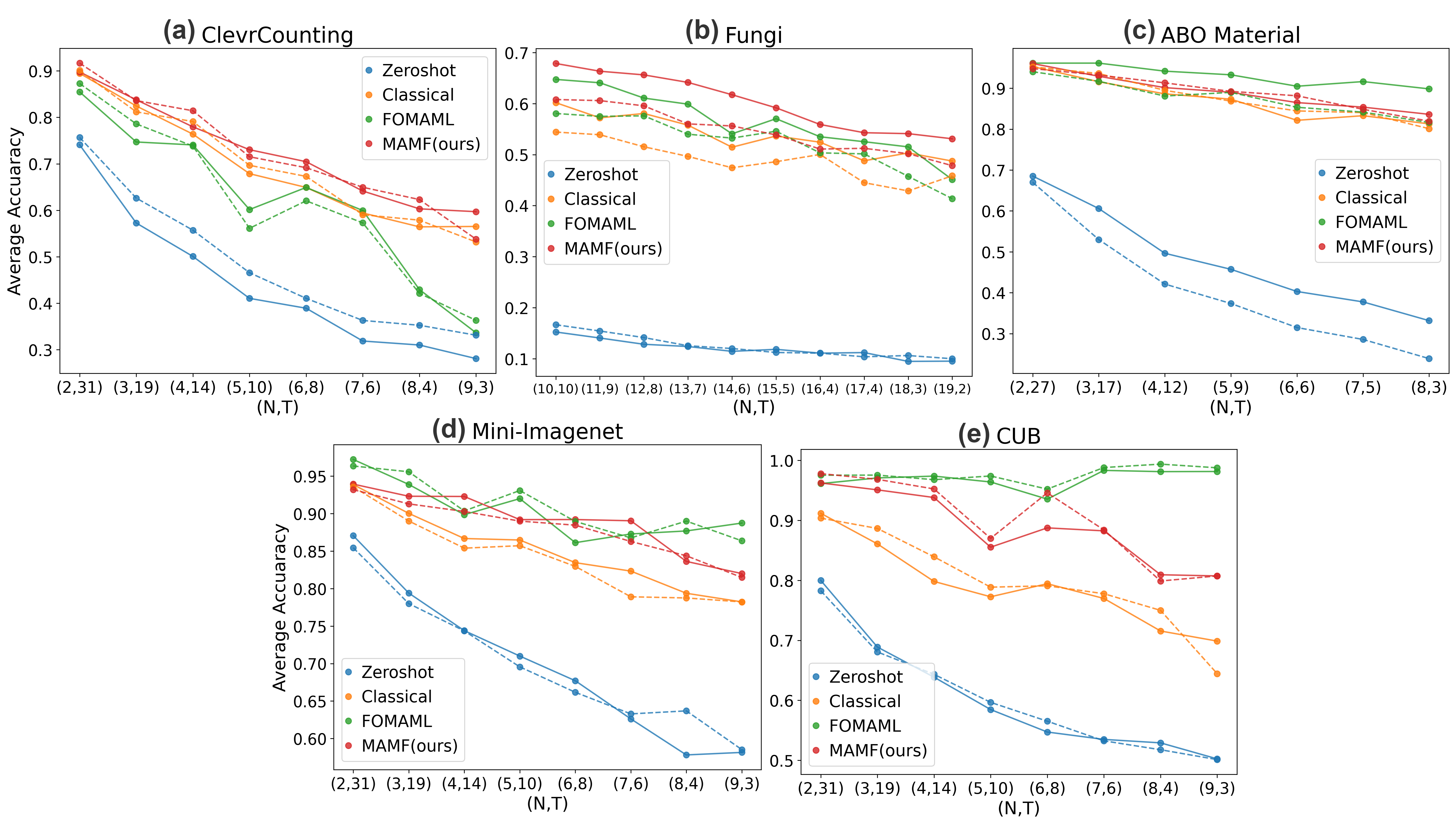}
  \caption{Average accuracy on development sets (\emph{dashed} line) and test sets (\emph{solid} line) of five datasets. 
  The x-axis shows the task configurations where $(N,T)$ refers to sampling $T$ tasks for $N$-way classification.  \emph{Zeroshot} refers to zero-shot CLIP without any fine-tuning during either training or {meta-testing}. \emph{Classical} refers to classical fine-tuning where we perform joint training on the entire training set. Both \emph{FOMAML} and \emph{MAMF} sample $N$-way $T$ tasks during training. \emph{MAMF} consistently outperforms \emph{Classical} on all datasets. 
  }
  \label{fig:result1}
\end{figure*}
\subsection{Results}
\textbf{Answer to Q1: Uniform task sampling is important.} 
As depicted in Figure~\ref{fig:result1}, 
comparing the performance of MAMF (red line) and Classical Fine-tuning (yellow line), MAMF consistently outperforms Classical Fine-tuning on all five datasets. 
Recall that the only difference between MAMF and Classical Fine-tuning is whether they perform uniform task sampling during training. This empirical result shows that task sampling itself serves as an important procedure for learning new concepts in a few-shot setting, even if with its simplest form, i.e. uniform sampling. 

\begin{figure}[th]
  \centering
  \includegraphics[width=0.44\textwidth]{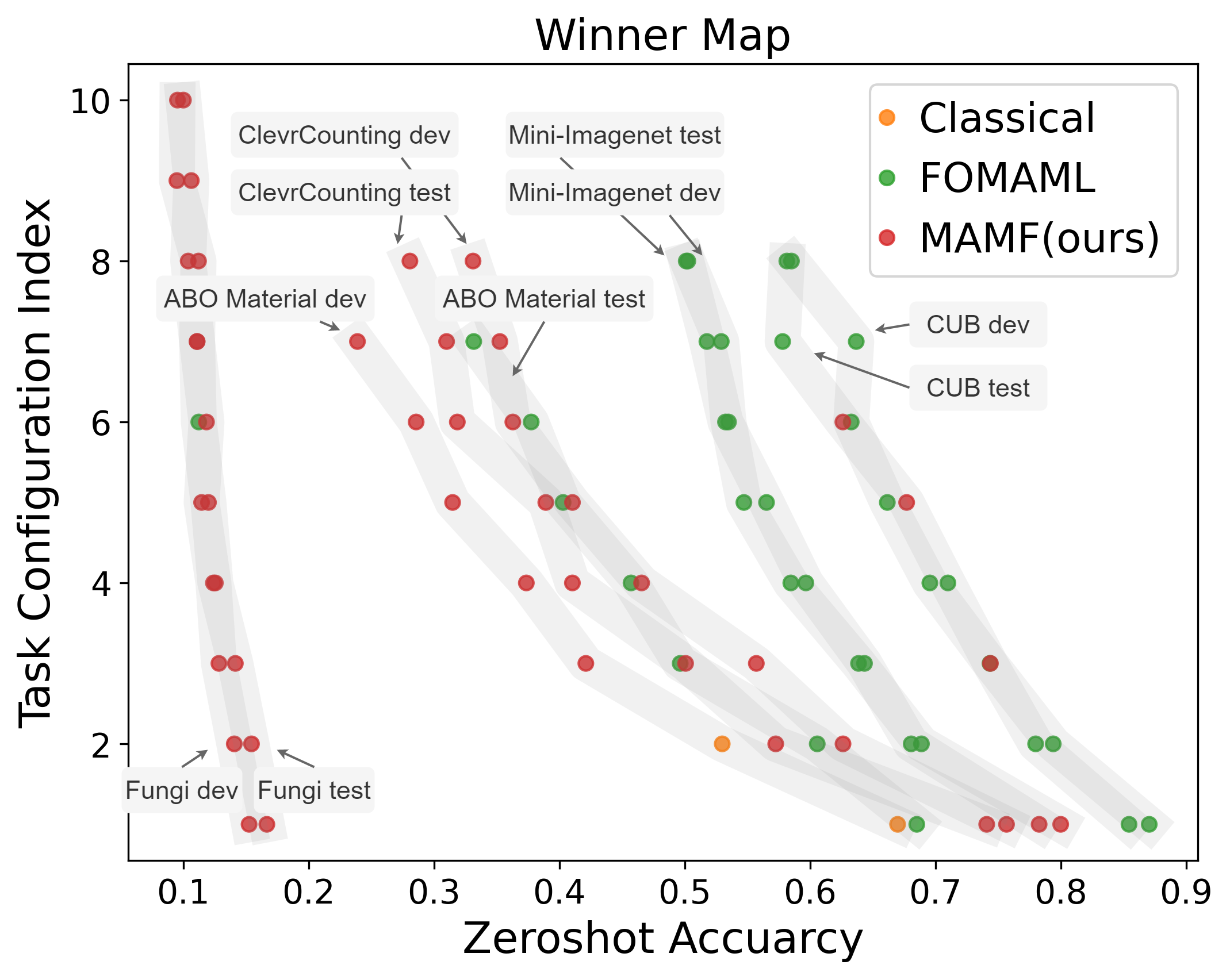}
  \caption{ Each thick shaded line represents a dataset split, e.g., test set of ClevrCounting. Each dot corresponds to one task configuration in Figure~\ref{fig:result1} such as $(N=5,T=10)$. The color of a dot represents the best-performing algorithm. \emph{MAMF} tends to outperform other algorithms when the problem is initially more challenging, i.e., when zero-shot accuracy is lower. }
  \label{fig:winner_map}
\end{figure}

%
\noindent
\textbf{Answer to Q2: MAML is not effective on learning initially challenging problems.}
One unexpected observation from Figure~\ref{fig:result1} is that, although FOMAML has the same task sampling procedure and more sophisticated optimization method than MAMF, it is outperformed by MAMF on many tasks.
We find that the effectiveness of FOMAML is highly sensitive to the zero-shot performance of the target task. Whenever the task is initially more challenging, i.e., with lower zero-shot performance, FOMAML tends to be less effective.
For example, on CUB (Figure~\ref{fig:result1} e) where the zero-shot accuracy ranges from $0.5$ to $0.8$, FOMAML outperforms other algorithms in most cases. However, on ClevrCounting (Figure~\ref{fig:result1} a) where the zero-shot accuracy ranges from $0.3$ to $0.75$, MAMF and even Classical Fine-tuning consistently outperform FOMAML. 
To further visualize this correlation, 
we plot a \emph{Winner Map} (Figure~\ref{fig:winner_map}) which depicts the best-performing method for each task configuration on all datasets.
We can see a clear pattern showing that FOMAML is only effective when the zero-shot performance is already high, while MAMF dominates on initially more challenging tasks. 

%% file: sections/conclusion.tex
\section{Conclusion}
In this paper, 
We demonstrate the importance of \textit{task sampling} by proposing a simple yet effective fine-tuning method MAMF. 
We further show novel insights on the limited effectiveness of the bi-level optimization. We hope our work encourages more research on improving few-shot transfer learning via better task sampling beyond uniform sampling. 

%% file: sections/appendix.tex
\section{Dataset Details}
\label{sec:appendix:dataset}
In this work, we compare few-shot image classification performance on five datasets representing various concepts including: ClevrCounting~\cite{johnson2017clevr}, Amazon Berkeley Objects (ABO)~\cite{collins2021abo} Material, Fungi~\cite{su2021realistic}, Mini-Imagenet~\cite{vinyals2016matching}, Caltech-UCSD Birds 200 (CUB)~\cite{WelinderEtal2010}. We randomly split each dataset into disjoint training, development, and test sets, and perform subsampling to frame the experiments in a few-shot setting. 
Specifically, for ABO Material, we construct a subset of the original dataset by clustering images according to their Material attribute. We then manually filter out noisy samples that have multiple major materials. Table~\ref{tab:datasets} shows the statistics of each dataset. 

We selectively add data augmentation\footnote{\url{https://pytorch.org/vision/stable/transforms.html}} for different datasets. 
By default we use \textit{RandomResizedCrop},\textit{ RandomHorizontalFlip} and \textit{Normalize} for all our five datasets. We further add \textit{ColorJitter} for Mini-Imagenet and ClevrCounting. We \textbf{disable} \textit{ColorJitter} for CUB, Fungi, and ABO Material since the color feature is essential for doing classification on these datasets. Following the original CLIP paper~\cite{pmlr-v139-radford21a}, the input images are resized to 224$\times$224.

\begin{table}[htb]
  \centering
  \small
  \begin{tabular}{cccccc}
    \toprule
    Dataset  & $M$    &  $S^{tr}$ & $S^{ts}_A$ & $S^{ts}_B$ \\
    \midrule
    ClevrCounting & 10 & 60 & 10 & 10 \\
    Fungi & 20 & 60 & 10 & 10 \\
    ABO Material & 9 & 50 & 15 & 15  \\
    Mini Imagenet & 10 & 60 & 10 & 10  \\
    CUB & 10 & 60 & 10 & 10  \\
    \bottomrule
  \end{tabular}
  \caption{Dataset statistics. $M$ is the total number of classes; $S^{tr}$ is the number of training samples per class; $S^{ts}_A$ and $S^{ts}_B$ are the number of support set and query set samples per class during \emph{meta-testing} respectively.}
\label{tab:datasets}
\end{table}

\begin{figure}[th]
    \centering
    \includegraphics[width=0.45 \textwidth]{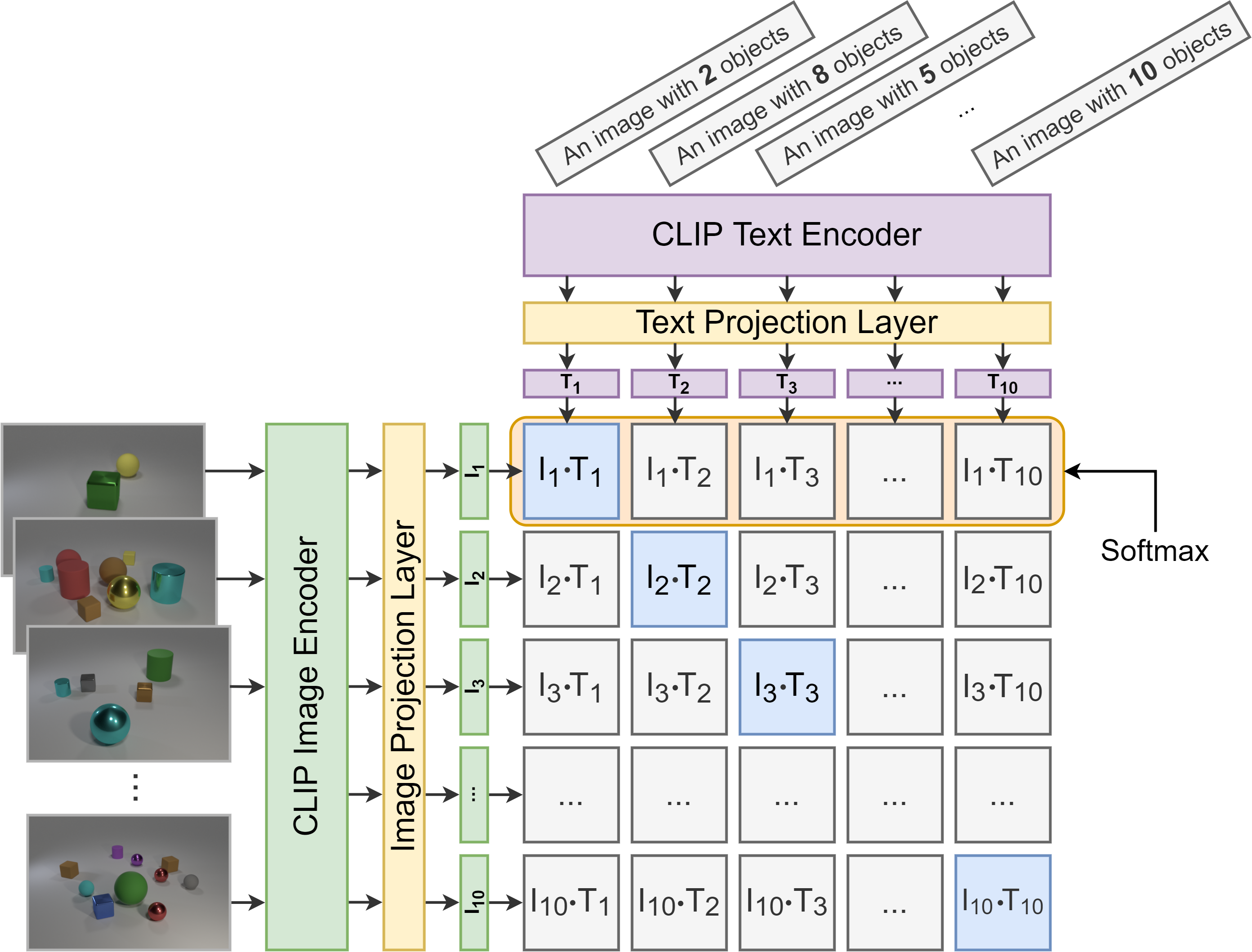}
    \caption{An illustration of the contrastive classification framework. We show a 10-way classification task on the Clevrcounting dataset. Each entry in the matrix is the similarity score (dot product) of an image embedding $\textbf{I}$ and a text embedding $\textbf{T}$. 
    }
    \label{fig:contrast}
\end{figure}

\section{Contrastive Image Classification Framework}
\label{sec:appendix:classification_framework}
We compare three algorithms (Classical Fine-tuning, MAML Fine-tuning, and MAMF) using an a contrastive classification framework based on pretrained CLIP~\cite{pmlr-v139-radford21a}. Instead of using a linear output layer mapping to $N$ logits corresponding to $N$ class labels, we directly compute the similarity between candidate text embeddings representing each class with the image embedding.
Specifically, we create the text representation for each class by using \textit{template prompts} filled with label names. A full list of templates we use for each dataset can be found in Table \ref{tab:template}.
Figure~\ref{fig:contrast} shows an example task from the ClevrCounting dataset, where each class is represented as a string such as ``An image with \textbf{2} objects". 
We then compute the dot product of each $<$image, text$>$ embedding pairs. For each row, the label with the highest similarity score is selected as the final prediction.

\begin{table}[th]
  \centering
  \small
  \begin{tabular}{cc}
    \toprule
    Dataset  & Text Input Template Example \\
    \midrule
    ClevrCounting & \text{An image of $<$\textbf{8}$>$ objects.}\\
    Fungi &  \text{A photo of $<$\textbf{mycena pura}$>$.}\\
    ABO Material &  \text{An image of a product made of $<$\textbf{glass}$>$}\\
    Mini Imagenet &  \text{A photo of $<$\textbf{walker hound}$>$.} \\
    CUB &  \text{A photo of $<$\textbf{baltimore oriole}$>$.}\\
    \bottomrule
  \end{tabular}
  \caption{Example templates with filled labels for all five datasets.}
\label{tab:template}
\end{table}

\section{Implementation Details}
\label{sec:appendix:implementation_details}
We use the pretrained CLIP\footnote{\url{https://huggingface.co/openai/clip-vit-base-patch32}}\cite{pmlr-v139-radford21a} with a ViT-B/32 Vision Transformer as image encoder and a masked self-attention Transformer as text encoder. The image embedding size is $768$ and the text embedding size is $512$. During training, we take the pre-projection image/text representation from the pretrained image/text encoder and feed them into a newly initialized\footnote{We use the Kaiming initialization implemented by Pytorch: \url{https://pytorch.org/cppdocs/api/function_namespacetorch_1_1nn_1_1init_1ac8a913c051976a3f41f20df7d6126e57.html}} image/text projection layer. We choose the pre-projection representation as prior work \cite{chen2020simple} has shown that in such contrastive models the hidden layer before the last projection head serves as a better representation. Finally, we obtain an image embedding and a text embedding with the same size of 512. Note that for the Zeroshot baseline, we use the original projection layer and directly test on the query set in \textit{meta-testing} without any fine-tuning. 
We train the model using cross-entropy loss for all three algorithms. 
We use the Adam optimizer~\cite{DBLP:journals/corr/KingmaB14} with learning rate $1e-6$ during training and $1e-7$ during \textit{meta-testing}. No weight decay is used for all algorithms during training and \textit{meta-testing}. We use the MAML wrapper from learn2learn\footnote{\url{https://github.com/learnables/learn2learn}}\cite{Arnold2020-ss} for training using first-order MAML.

\section{Detailed Results}
\label{sec:appendix:detailed_results}
Table~\ref{tab:all_results_dev} shows the detailed accuracy and standard deviation on the development sets and test sets of all the datasets shown in Figure 2 in the main paper. The $(N,T)$ column represents the task configurations, where $N$ stands for an $N$-way classification task and $T$ stands for the total number of sampled tasks. 
Since the tasks are randomly sampled from the class distribution, in order to cover all classes with high probability during testing, we set the number of sampled tasks to be: $T = \frac{log(0.001)}{log(1-\frac{N}{M})}$, where $M$ is the total number of classes. That is, with probability higher than $0.999$, we can cover all classes if sampling $T$ tasks. 
Columns with name \textit{Zeroshot}, \textit{Classical}, \textit{MAMF}, and \textit{FOMAML} represent models using Zeroshot CLIP, Classical Fine-tuning, Model-Agnostic Multitask Fine-tuning and first-order MAML respectively. The superscript on each accuracy percentage number indicates standard deviation across five random runs.

{
\setlength{\tabcolsep}{0.2em} 
\begin{table*}[htb]
\small
\centering
\caption{\small
Detailed average accuracy (\%) and standard deviation on the development set and test set of all five datasets. The (N, T) column represents the task configurations consistent with the x-axis in Figure 3 in the main paper. Note that for the ABO Material dataset, we have 9 classes in total, so a task has up to 8-way classification. And for the Fungi dataset, which has 20 classes in total, we test on 10-way to 19-way classification tasks.}
\vspace{-3mm}
\begin{tabular}{c c c c c c c c c c c c}
\toprule
\multicolumn{1}{c}{Dataset} & \multicolumn{1}{c}{(N, T)} & Zeroshot & Classical & MAMF & FOMAML & \multicolumn{1}{c}{Dataset} & (N, T) & Zeroshot & Classical & MAMF & FOMAML\\
\toprule
{} & (2, 27) & \report{68.5}{2.6} & \report{95.3}{1.8} & \report{96.0}{1.8} & \report{\textbf{96.1}}{2.4} & {} & (2, 27) & \report{67.0}{2.6} & \report{95.2}{0.6} & \report{\textbf{94.8}}{1.2} & \report{94.0}{0.4}\\
{} & (3, 17) & \report{60.6}{6.0} & \report{91.6}{3.0} & \report{92.9}{2.4} & \report{\textbf{96.1}}{1.0} & {} & (3, 17) & \report{53.0}{1.9} & \report{\textbf{93.6}}{0.9} & \report{93.2}{0.9} & \report{91.7}{1.3}\\
{ABO} & (4, 12) & \report{49.6}{3.4} & \report{88.6}{0.6} & \report{90.1}{0.7} & \report{\textbf{94.2}}{1.6} & {ABO} & (4, 12) & \report{42.1}{2.8} & \report{89.5}{2.5} & \report{\textbf{91.3}}{2.2} & \report{88.1}{1.7}\\
{Material} & (5, 9) & \report{45.7}{1.6} & \report{87.3}{1.5} & \report{89.0}{1.2} & \report{\textbf{93.3}}{0.7} & {Material} & (5, 9) & \report{37.4}{2.6} & \report{86.8}{2.2} & \report{\textbf{89.2}}{1.4} & \report{89.0}{1.4}\\
{Test} & (6, 6) & \report{40.3}{2.1} & \report{82.1}{2.1} & \report{86.5}{2.5} & \report{\textbf{90.5}}{1.4} & {Dev} & (6, 6) & \report{31.5}{1.8} & \report{84.4}{1.6} & \report{\textbf{88.1}}{3.2} & \report{85.3}{2.9}\\
{} & (7, 5) & \report{37.8}{3.4} & \report{83.3}{3.0} & \report{85.4}{4.0} & \report{\textbf{91.6}}{1.7} & {} & (7, 5) & \report{28.6}{2.0} & \report{84.0}{1.5} & \report{\textbf{84.8}}{1.2} & \report{84.2}{3.1}\\
{} & (8, 3) & \report{33.2}{0.7} & \report{81.3}{0.8} & \report{83.6}{1.6} & \report{\textbf{89.8}}{0.9} & {} & (8, 3) & \report{23.9}{2.7} & \report{80.1}{1.3} & \report{\textbf{81.9}}{1.8} & \report{81.6}{1.2}\\
\toprule
{} & (2, 31) & \report{74.1}{2.2} & \report{89.5}{1.7} & \report{\textbf{89.8}}{1.5} & \report{85.5}{2.2} & {} & (2, 31) & \report{75.7}{3.4} & \report{90.1}{2.7} & \report{\textbf{91.7}}{2.5} & \report{87.3}{1.4}\\
{} & (3, 19) & \report{57.3}{2.1} & \report{82.5}{3.3} & \report{\textbf{83.8}}{3.8} & \report{74.7}{4.5} & {} & (3, 19) & \report{62.6}{1.5} & \report{81.2}{2.5} & \report{\textbf{83.6}}{3.0} & \report{78.6}{4.5}\\
{Clevr-} & (4, 14) & \report{50.1}{2.0} & \report{76.4}{2.0} & \report{\textbf{78.0}}{3.3} & \report{74.1}{1.6} & {Clevr-} & (4, 14) & \report{55.7}{1.9} & \report{79.1}{2.0} & \report{\textbf{81.4}}{1.1} & \report{73.8}{5.3}\\
{Counting} & (5, 10) & \report{41.0}{2.5} & \report{67.9}{3.0} & \report{\textbf{73.0}}{2.8} & \report{60.2}{9.5} & {Counting} & (5, 10) & \report{46.6}{3.7} & \report{69.7}{4.6} & \report{\textbf{71.5}}{6.1} & \report{56.1}{0.9}\\
{Test} & (6, 8) & \report{38.9}{2.6} & \report{64.9}{3.9} & \report{\textbf{70.5}}{3.0} & \report{64.9}{5.0} & {Dev} & (6, 8) & \report{41.0}{1.3} & \report{67.3}{2.6} & \report{\textbf{69.2}}{5.2} & \report{62.0}{3.6}\\
{} & (7, 6) & \report{31.9}{0.9} & \report{59.4}{3.3} & \report{\textbf{64.1}}{1.7} & \report{60.0}{6.9} & {} & (7, 6) & \report{36.3}{1.9} & \report{59.0}{4.8} & \report{\textbf{65.0}}{4.2} & \report{57.3}{1.9}\\
{} & (8, 4) & \report{31.0}{1.3} & \report{56.4}{5.8} & \report{\textbf{60.3}}{3.2} & \report{42.9}{5.6} & {} & (8, 4) & \report{35.3}{1.2} & \report{57.9}{4.6} & \report{\textbf{62.3}}{3.5} & \report{42.1}{9.2}\\
{} & (9, 3) & \report{28.1}{1.2} & \report{56.5}{3.1} & \report{\textbf{59.7}}{4.8} & \report{33.6}{13.2} & {} & (9, 3) & \report{33.1}{2.3} & \report{53.2}{3.3} & \report{\textbf{53.8}}{2.6} & \report{36.3}{12.6}\\
\toprule
{} & (2, 31) & \report{80.0}{2.6} & \report{91.2}{1.6} & \report{\textbf{96.2}}{1.8} & \report{96.1}{2.9} & {} & (2, 31) & \report{78.3}{1.1} & \report{90.4}{1.9} & \report{\textbf{97.8}}{1.5} & \report{97.5}{1.4}\\
{} & (3, 19) & \report{68.9}{1.6} & \report{86.1}{5.6} & \report{95.1}{0.9} & \report{\textbf{97.1}}{3.3} & {} & (3, 19) & \report{68.1}{2.3} & \report{88.7}{7.1} & \report{96.8}{2.3} & \report{\textbf{97.5}}{2.5}\\
{} & (4, 14) & \report{63.9}{3.3} & \report{79.8}{4.4} & \report{93.8}{2.6} & \report{\textbf{97.4}}{1.9} & {} & (4, 14) & \report{64.3}{1.9} & \report{83.9}{4.6} & \report{95.2}{2.8} & \report{\textbf{96.8}}{2.8}\\
{CUB} & (5, 10) & \report{58.5}{2.6} & \report{77.3}{3.3} & \report{85.5}{6.6} & \report{\textbf{96.4}}{1.4} & {CUB} & (5, 10) & \report{59.7}{2.3} & \report{78.9}{5.4} & \report{87.0}{6.2} & \report{\textbf{97.4}}{2.2}\\
{Test} & (6, 8) & \report{54.7}{2.0} & \report{79.4}{5.5} & \report{88.8}{3.0} & \report{\textbf{93.5}}{5.6} & {Dev} & (6, 8) & \report{56.5}{1.3} & \report{79.1}{0.6} & \report{94.6}{3.1} & \report{\textbf{95.2}}{5.2}\\
{} & (7, 6) & \report{53.5}{1.9} & \report{77.0}{7.9} & \report{88.3}{3.5} & \report{\textbf{98.3}}{0.3} & {} & (7, 6) & \report{53.3}{2.0} & \report{77.8}{3.4} & \report{88.4}{3.9} & \report{98.8}{0.0}\\
{} & (8, 4) & \report{52.9}{2.6} & \report{71.6}{7.4} & \report{80.9}{3.8} & \report{\textbf{98.1}}{0.5} & {} & (8, 4) & \report{51.8}{1.7} & \report{75.0}{7.2} & \report{79.9}{4.7} & \report{99.4}{0.4}\\
{} & (9, 3) & \report{50.2}{1.3} & \report{69.9}{5.1} & \report{80.7}{4.1} & \report{\textbf{98.1}}{0.7} & {} & (9, 3) & \report{50.1}{2.5} & \report{64.4}{8.3} & \report{80.7}{3.3} & \report{\textbf{98.8}}{0.9}\\
\toprule
{} & (2, 31) & \report{87.1}{1.4} & \report{93.9}{1.5} & \report{93.9}{1.5} & \report{\textbf{97.2}}{1.4} & {} & (2, 31) & \report{85.5}{2.7} & \report{93.6}{0.7} & \report{93.2}{3.0} & \report{\textbf{96.4}}{0.4}\\
{} & (3, 19) & \report{79.4}{2.3} & \report{90.0}{1.6} & \report{92.3}{1.5} & \report{\textbf{93.9}}{2.6} & {} & (3, 19) & \report{78.0}{3.4} & \report{89.0}{1.5} & \report{91.3}{1.5} & \report{\textbf{95.6}}{0.8}\\
{Mini} & (4, 14) & \report{74.4}{3.1} & \report{86.7}{1.4} & \report{\textbf{92.3}}{1.4} & \report{89.9}{5.1} & {Mini} & (4, 14) & \report{74.4}{4.2} & \report{85.4}{2.4} & \report{90.3}{0.7} & \report{\textbf{90.4}}{5.7}\\
{ImageNet} & (5, 10) & \report{71.0}{2.4} & \report{86.5}{0.7} & \report{\textbf{89.2}}{1.4} & \report{92.0}{0.7} & {ImageNet} & (5, 10) & \report{69.6}{5.7} & \report{85.7}{3.3} & \report{89.0}{1.2} & \report{\textbf{93.1}}{2.6}\\
{Test} & (6, 8) & \report{67.7}{3.4} & \report{83.5}{1.1} & \report{\textbf{89.2}}{1.6} & \report{86.1}{2.8} & {Dev} & (6, 8) & \report{66.2}{2.8} & \report{83.0}{1.1} & \report{88.5}{1.5} & \report{\textbf{89.0}}{2.2}\\
{} & (7, 6) & \report{62.6}{3.0} & \report{82.3}{1.5} & \report{\textbf{89.0}}{1.3} & \report{87.3}{3.1} & {} & (7, 6) & \report{63.3}{2.3} & \report{78.9}{2.6} & \report{86.3}{0.8} & \report{\textbf{86.8}}{2.9}\\
{} & (8, 4) & \report{57.8}{1.9} & \report{79.4}{2.9} & \report{83.6}{3.4} & \report{\textbf{87.7}}{6.0} & {} & (8, 4) & \report{63.7}{3.2} & \report{78.7}{5.0} & \report{84.4}{1.7} & \report{\textbf{89.0}}{1.0}\\
{} & (9, 3) & \report{58.1}{2.6} & \report{78.2}{3.1} & \report{82.0}{2.9} & \report{\textbf{88.7}}{1.5} & {} & (9, 3) & \report{58.5}{2.8} & \report{78.2}{1.9} & \report{81.5}{1.9} & \report{\textbf{86.4}}{3.1}\\
\toprule
{} & (10, 8) & \report{15.2}{0.8} & \report{60.2}{1.5} & \report{\textbf{67.9}}{2.6} & \report{64.7}{2.7} & {} & (10, 8) & \report{16.7}{1.4} & \report{54.4}{1.1} & \report{\textbf{60.8}}{2.9} & \report{58.1}{1.5}\\
{} & (11, 8) & \report{14.1}{0.5} & \report{57.2}{2.2} & \report{\textbf{66.3}}{1.7} & \report{64.1}{1.4} & {} & (11, 8) & \report{15.4}{1.2} & \report{53.9}{2.5} & \report{\textbf{60.6}}{1.0} & \report{57.5}{1.5}\\
{} & (12, 8) & \report{12.8}{0.7} & \report{58.1}{1.9} & \report{\textbf{65.6}}{2.6} & \report{61.1}{2.6} & {} & (12, 8) & \report{14.1}{0.8} & \report{51.5}{3.9} & \report{\textbf{59.6}}{2.6} & \report{57.6}{2.7}\\
{} & (13, 7) & \report{12.4}{0.7} & \report{55.8}{1.1} & \report{\textbf{64.2}}{2.7} & \report{59.9}{2.3} & {} & (13, 7) & \report{12.5}{1.0} & \report{49.6}{4.1} & \report{\textbf{56.0}}{1.9} & \report{54.0}{1.6}\\
{Fungi} & (14, 6) & \report{11.5}{1.2} & \report{51.5}{2.6} & \report{\textbf{61.7}}{2.2} & \report{54.1}{4.3} & {Fungi} & (14, 6) & \report{12.0}{0.6} & \report{47.4}{3.3} & \report{\textbf{55.6}}{3.4} & \report{53.2}{2.6}\\
{Test} & (15, 5) & \report{11.8}{0.8} & \report{53.7}{1.7} & \report{\textbf{59.2}}{3.8} & \report{57.0}{1.3} & {Dev} & (15, 5) & \report{11.2}{0.1} & \report{48.6}{0.9} & \report{53.9}{1.7} & \report{\textbf{54.6}}{2.8}\\
{} & (16, 4) & \report{11.1}{0.5} & \report{52.4}{2.6} & \report{\textbf{55.9}}{1.5} & \report{53.5}{2.5} & {} & (16, 4) & \report{11.1}{0.6} & \report{50.0}{3.0} & \report{\textbf{51.1}}{2.8} & \report{50.3}{4.4}\\
{} & (17, 4) & \report{11.2}{0.5} & \report{48.8}{3.1} & \report{\textbf{54.3}}{1.4} & \report{52.5}{1.0} & {} & (17, 4) & \report{10.4}{1.0} & \report{44.5}{1.4} & \report{\textbf{51.2}}{2.1} & \report{50.1}{1.7}\\
{} & (18, 3) & \report{9.5}{0.3} & \report{50.3}{3.0} & \report{\textbf{54.1}}{2.1} & \report{51.5}{2.4} & {} & (18, 3) & \report{10.6}{0.3} & \report{42.9}{3.1} & \report{\textbf{50.2}}{1.8} & \report{45.7}{2.0}\\
{} & (19, 2) & \report{9.5}{0.7} & \report{48.7}{3.4} & \report{\textbf{53.1}}{3.3} & \report{45.1}{2.1} & {} & (19, 2) & \report{10.0}{1.3} & \report{45.8}{3.2} & \report{\textbf{47.8}}{1.6} & \report{41.4}{3.0}\\
\toprule
\end{tabular}
\label{tab:all_results_dev}
\end{table*}
}